\documentclass[aps,prl,reprint,groupedaddress]{revtex4-2}
\usepackage{amsmath, amsfonts, amssymb, amsthm}
\usepackage{dcolumn}
\usepackage{color}

\begin{document}


\title{Entanglement Entropy as a Probe Beyond the Horizon}

\author{Konstantinos Boutivas}
\email{kboutivas@phys.uoa.gr}
\affiliation{National and Kapodistrian University of Athens, Department of Physics, 15784 Zografou, Attiki, Greece}
\author{Dimitrios Katsinis}
\email{dkatsinis@phys.uoa.gr}
\affiliation{National and Kapodistrian University of Athens, Department of Physics, 15784 Zografou, Attiki, Greece}
\author{Georgios Pastras}
\altaffiliation[Also at ]{Laboratory for Manufacturing Systems and Automation, Department of Mechanical Engineering and Aeronautics, University of Patras, 26110 Patra, Greece}
\email{pastras@lms.mech.upatras.gr}
\affiliation{National and Kapodistrian University of Athens, Department of Physics, 15784 Zografou, Attiki, Greece}
\author{Nikolaos Tetradis}
\email{ntetrad@phys.uoa.gr}
\affiliation{National and Kapodistrian University of Athens, Department of Physics, 15784 Zografou, Attiki, Greece}



\date{\today}

\begin{abstract}
The entanglement entropy of a free field in de Sitter space is enhanced by the squeezing of its modes. We show analytically that the expansion induces a term in the entanglement entropy  that depends logarithmically on the size of the overall system, which may extend beyond the horizon. In cosmology the size of the system can be identified with the size of a spatially finite universe, or with the wavelength of the first mode that exited the horizon in the beginning of inflation. 
\end{abstract}


\maketitle

\section{Introduction}

According to the standard inflationary paradigm, the quantum fluctuations of massless fields become classical upon horizon exit. Upon re-entry after the end of inflation, they result in density perturbations that generate the observed large-scale structure of the universe \cite{Mukhanov:1981xt,Hawking:1982cz,Guth:1982ec,Starobinsky:1982ee,Bardeen:1983qw}. This quantum-to-classical transition can be attributed to the fact that the field mode functions consist of a growing and a decaying term, with the growing one becoming dominant after horizon exit. If the decaying term is neglected, the field commutes with its conjugate momentum. Therefore, it can be treated as a classical stochastic field. This simple line of reasoning does not take into account the fact that, no matter how suppressed the decaying mode is, the field and its conjugate momentum always obey the canonical commutation relation. Strictly speaking, the field never loses its quantum nature, even though measuring quantum effects may be extremely difficult \cite{Albrecht:1992kf,Polarski:1995jg,Kiefer:1998qe,Kiefer:1998pb,Allen:1999xw,Maldacena:2015bha}.

It is known that the field modes evolve from a simple harmonic oscillator ground state to an increasingly squeezed one, which ceases to be a state of minimal uncertainty \cite{Albrecht:1992kf,Grishchuk:1990bj}. From this point of view, it seems possible that the quantum nature of the fluctuations may be discernible, as long as one looks at the appropriate quantity. There is a connection between the squeezing of the field modes in a cosmological setting and the production of entropy \cite{Brandenberger:1992jh,Brandenberger:1992sr,Prokopec:1992ia,Matacz:1992tp,Gasperini:1992xv,Kiefer:1999sj}. We focus here on entanglement entropy, which is a quantity that reflects directly the quantum nature of the fluctuations. We work in coordinate space, in an expanding background that includes a cosmological horizon. We compute the entanglement entropy of a subsystem enclosed by an entangling surface. We show that the expansion enhances the effects of entanglement, leading to an increase in entropy, and also inducing its dependence on global properties of space. For a de Sitter background, the leading correction to the entanglement entropy beyond its flat-space value depends on the size of the \emph{entire system}, which may extend beyond the horizon.

The archetypical calculations in $(3+1)$-dimensional flat space \cite{Srednicki:1993im,Bombelli:1986rw} led to the unexpected finding that the leading contribution to the entanglement entropy of a free massless scalar field is proportional to the area of the surface enclosing the subsystem under consideration. This result is attributed to the strong entanglement of short-distance modes on either side of the entangling surface, which also makes the entropy UV-divergent. The numerical calculation of the coefficient of the area-law term is a relatively easy task nowadays, given the available computing power. However, the computation of the coefficients of the subleading terms, such as the universal logarithmic term, requires an elaborate numerical study \cite{Lohmayer:2009sq}. Analytical calculations, even for a free field theory, are more difficult \cite{Casini:2009sr}. There are only a few analytical results, mainly in lower dimensions \cite{Callan:1994py,Holzhey:1994we,Korepin:2004zz,Calabrese:2004eu,Calabrese:2009qy}. The effects of the expansion have been analyzed even less \cite{Maldacena:2012xp,Kanno:2014lma,Iizuka:2014rua,Kanno:2016qcc,Berges:2017hne,Boutivas:2023ksg,Boutivas:2023mfg}.

We base our analysis on the approach of Srednicki in its seminal work \cite{Srednicki:1993im}. We calculate the entanglement entropy of a massless scalar field in the Bunch-Davies vacuum of $(3+1)$-dimensional de Sitter space. The necessary formalism has been developed in our recent work \cite{Boutivas:2023ksg,Boutivas:2023mfg} that uses the generalization of Srednicki's method for harmonic systems at squeezed states \cite{Katsinis:2023hqn,Katsinis:2024sek}.  We consider a spherical entangling surface, expand the field in spherical moments, and discretize the radial direction. The theory is decomposed into an infinite number of independent sectors, each of them corresponding to a fixed value of angular momentum $\ell$. Each sector amounts to a discretized effective $(1+1)$-dimensional system, whose size is set by the number of points $N$ of the radial lattice. The entanglement entropy is calculated as the sum of the contributions of all sectors, namely
\begin{equation}
S=\sum_{\ell=0}^\infty\sum_{m=-\ell}^{\ell}S_{\ell m}=\sum_{\ell=0}^\infty\left(2\ell+1\right)S_{\ell},
\end{equation}
taking into account that the dynamics of each sector depends only on $\ell$.

A crucial step in the calculation is taking the limit $N\sim L/\epsilon \to \infty$, where $L$ is the size of the system and $\epsilon$ the UV cutoff, identified with the lattice spacing. The main obstacle is that subleading terms in the entropy must be disentangled from finite-size effects that inevitably appear in the calculation. This problem is ubiquitous in numerical calculations of the entropy. A method for dealing with the finite-size effects in flat space has been developed in \cite{Lohmayer:2009sq}. An important observation is that these effects are suppressed for large angular momentum. The higher the angular momentum, the more disentangled the local degrees of freedom of the theory are. This happens because the angular momentum acts as a position-dependent mass, making the normal modes more and more localized around certain points as its value increases. In $(3+1)$-dimensional flat space it turns out that the finite-size effects scale as \cite{Lohmayer:2009sq}
\begin{equation}
S^{\textrm{flat}}_{\textrm{fs}}\propto N^{-2\left(\ell+1\right)}.
\label{finitesizeflat}
\end{equation}
It is evident that these terms vanish in the infinite-size limit.

In an accompanying publication \cite{Boutivas:2024aq}, we perform a thorough numerical study of entanglement entropy in de Sitter space, modifying appropriately the methodology developed for flat space in \cite{Lohmayer:2009sq}. We consider a field in the Bunch-Davies vacuum and observe how entanglement is enhanced through the squeezing of the state. It turns out that the dominant finite-size effects now scale as
\begin{equation}
S^{\textrm{dS}}_{\textrm{fs}}\propto\begin{cases}\ln N, & \ell=0,\\
N^{-2\ell},& \ell\neq 0.
\end{cases}
\label{finitesizedS}
\end{equation}
Rather surprisingly, the contribution of the $\ell=0$ sector does not vanish in the infinite-size limit. The numerical calculation indicates that the resulting contribution to the entropy is of the novel form
\begin{equation}\label{eq:S_IR}
S^{\textrm{dS}}_{\textrm{IR}}\propto \frac{R^2}{\tau^2}\ln\frac{L}{\epsilon},
\end{equation}
where $L$ is the radius of the \emph{overall system}, $R$ the radius of the subsystem, $\tau$ the conformal time and $\epsilon$ the UV cutoff. The coefficient of this term is found numerically to be approximately equal to $1/3$. In this work we verify this fact with an analytical calculation.

The means to perform this calculation are provided by our recent work \cite{Katsinis:2024gef}, in which the $(1+1)$-dimensional discretized theory is studied analytically in order to obtain the continuum limit of the entanglement spectrum, the entanglement entropy and the corresponding modular Hamiltonian. The main idea is to take the continuum limit of the eigenvalue equations of the discrete problem, in which matrices become kernels and matrix multiplication turns into integration. As we discussed above, the $\ln N$ term stems solely from the  sector of the theory with vanishing angular momentum. This sector is identical to the $(1+1)$-dimensional free scalar field theory, so that the methodology and  some results of \cite{Katsinis:2024gef} are directly applicable to our calculation here.

\section{The Discretized Theory }

We consider a free massless scalar field $\phi \left( \tau , \mathbf{x} \right)$ in $(3+1)$-dimensional de Sitter space. Performing the field redefinition $\phi \left( \tau , \mathbf{x} \right) = f \left( \tau , \mathbf{x} \right) / a \left( \tau \right)$, where the scale factor is $a(\tau)=-1/(H\tau)$ and $H$ is the Hubble constant, yields the action
\begin{equation}
\mathcal{S} = \frac{1}{2} \int d\tau \, d^3 \mathbf{x} \, \left( \dot{f}^2 - \left( \nabla f \right)^2 + \frac{\ddot{a}}{a} f^2 \right),
\label{eq:action_dS}
\end{equation}
with a canonical kinetic term. The field $f \left( \tau , \mathbf{x} \right)$ obtains a time-dependent squared mass  $-\ddot{a}/a=-2/\tau^2$. We consider a spherical region of finite radius $L$ and expand the field in its spherical moments $f_{\ell m}(r)$. The Hamiltonian corresponding to each $(\ell,m)$-sector of the theory reads \cite{Huerta:2022tpq}
\begin{multline}\label{eq:Effective_H_2d}
H_{\ell m}=\frac{1}{2}\int_0^L dr \bigg[\pi_{\ell m}^2+\left(\partial_r f_{\ell m}\right)^2 \\ +\left(\frac{\ell\left(\ell+1\right)}{r^2}-\frac{2}{\tau^2}\right)f_{\ell m}^2\bigg]-\frac{1}{2}\frac{f_{\ell m}^2}{r}\bigg\vert^L_{0}.
\end{multline}
Considering Dirichlet boundary conditions we can drop the boundary term. Then we discretize the radial coordinate as $r=j\epsilon$, with $L=(N+1)\epsilon$, obtaining
\begin{multline}
H_{\ell m} = \frac{1}{2 \epsilon} \sum\limits_{j = 1}^N \bigg[ \pi_{\ell m,j}^2 + \left(f_{\ell m,j + 1} - f_{\ell m,j}\right)^2 \\+ \left( \frac{\ell \left( {\ell + 1} \right)}{j^2} - \frac{2\epsilon^2}{\tau^2} \right) f_{\ell m,j}^2 \bigg] .
\label{eq:Hamiltonian_discretized_dS}
\end{multline}
The discretization we used here is slightly different, but completely equivalent, to the one used by Srednicki \cite{Srednicki:1993im}. The conformal time $\tau$ is naturally measured in units of $\epsilon$, which we set equal to 1.

The system lies in the Bunch-Davies vacuum, which implies that the wave functions of its modes for $\tau\to -\infty$ must reduce to those in flat space.  As a result, each mode of the system is described by a wave function \cite{Boutivas:2023ksg,Boutivas:2023mfg}
\begin{multline}
F(\tau,f)=\frac{1}{\sqrt{b(\tau)}}\,\exp\left({\frac{i}{2}\frac{b'(\tau)}{b(\tau)}f^2}\right)\\
\times F^0\left(\int \frac{d\tau}{b^2(\tau)},\frac{f}{b(\tau)}\right)
\label{solschrod} 
\end{multline}
where $F^0(\tau,f)$ is a solution of the standard simple harmonic oscillator with constant frequency $\omega_0$. The function $b(\tau)=\sqrt{1+1/(\omega^2_0 \tau^2)}$ quantifies the effect of squeezing. It appoaches 1 for $\tau\to-\infty$, while it diverges for $\tau\to 0^-$. For the ground state we obtain
\begin{multline}\label{groundstatee}
F \left( \tau , x \right)=\sqrt[4]{\frac{\omega_0}{\pi}\frac{1}{1+\frac{1}{\omega_0^2\tau^2}}}\exp \left[ - \frac{\omega_0}{2}\frac{\omega_0^2 \tau^2+\frac{i}{\omega_0 \tau}} {1 + \omega_0^2 \tau^2} x^2 \right]\\
\times\exp\left[-\frac{i}{2}\left(\omega_0\tau-\arctan\left(\omega_0\tau\right)-\frac{\pi}{2}\right)\right].
\end{multline}

We focus on the $\ell=0$ sector of the theory, because the effect we are interested in originates in this sector only. The wave function describing the degrees of freedom of this sector reads 
\begin{equation}\label{eq:wave_function}
\Psi \left( \mathbf{x} \right) \propto \exp \left( - \frac{1}{2} x^T W x \right),
\end{equation}
where the matrix $W$ has the form
\begin{equation}
W = \Omega \left[ I - \left( I - \frac{i}{ \tau}\Omega^{-1} \right) \left(I + \Omega^2 \tau^2\right)^{-1}  \right].
\end{equation}
The matrix $\Omega$ is the positive square root of the flat-space coupling matrix $K$, which can be read directly from the Hamiltonian \eqref{eq:Hamiltonian_discretized_dS}.
It is given by
\begin{equation}\label{eq:couplings_matrix}
K_{ij}=2\delta_{i,j}-\delta_{i,j+1}-\delta_{i+1,j}.
\end{equation}

We consider the subsystem $A$ consisting of the first $n$ oscillators, and we trace out the remaining $N-n$ ones that constitute subsystem $A^C$. Throughout the paper we denote the blocks of $N\times N$ matrices as $\Omega=\begin{pmatrix}
\Omega_A & \Omega_B\\
\Omega_B^T & \Omega_C
\end{pmatrix}$. In particular, the matrix $M_A$ is $n\times n$, whereas the matrix $M_C$ is $(N-n)\times(N-n)$. The entanglement entropy of Gaussian states, like the one in Eq. \eqref{eq:wave_function}, can be calculated by making use of correlation functions \cite{Sorkin:2012sn}. We define the $2N\times 2N$ matrix $\mathcal{M} = 2i J \mathrm{Re}\, \left( M \right)$, where
\begin{equation}
M = \begin{pmatrix}
\left\langle x_i x_j \right\rangle & \left\langle x_i \pi_j\right\rangle \\
\left\langle x_i \pi_j \right\rangle^T & \left\langle \pi_i \pi_j \right\rangle
\end{pmatrix} , \qquad J = \begin{pmatrix}
0 & I \\
-I & 0
\end{pmatrix} .
\end{equation} 
The $2n\times 2n$  matrix $\mathcal{M}$ for subsystem $A$ has a relatively simple form in terms of the blocks of half-integer powers of the couplings matrix $K$ \cite{Boutivas:2024aq}, namely
\begin{widetext}
\begin{equation}\label{eq:matrix_M_cal}
\mathcal{M}_A = i \begin{pmatrix}
\frac{1}{\tau^3} \left( \Omega^{-3} \right)_A & \left( \Omega \right)_A - \frac{1}{\tau^2} \left( \Omega^{-1} \right)_A + \frac{1}{\tau^4} \left( \Omega^{-3} \right)_A \\
- \left( \Omega^{-1} \right)_A - \frac{1}{\tau^2} \left( \Omega^{-3} \right)_A & -\frac{1}{\tau^3} \left( \Omega^{-3} \right)_A 
\end{pmatrix}.
\end{equation}
\end{widetext}
Its eigenvalues come in pairs $\pm \lambda_i$ and satisfy $\left| \lambda_i \right| \geq 1$. The entanglement entropy can be computed as
\begin{equation}\label{eq:SEE_of_M_cal}
S = \sum_{i=0}^{n} \left( \frac{\lambda_i + 1}{2} \ln \frac{\lambda_i + 1}{2} - \frac{\lambda_i - 1}{2} \ln \frac{\lambda_i - 1}{2} \right),
\end{equation}
where the sum is performed only over the positive eigenvalues.

Calculating analytically the eigenvalues of \eqref{eq:matrix_M_cal} for an arbitrary value of $\tau$ is extremely hard. However, it is possible to implement perturbation theory for large $\vert \tau\vert$ in order to calculate the leading corrections to the entanglement entropy of flat space. The details of the perturbative expansion are given in \cite{Boutivas:2024aq}. The calculation is facilitated if we consider the square of the matrix $\mathcal{M}_A$, along with a similarity transformation, obtaining
\begin{widetext}
\begin{equation}\label{eq:M_cal_sq}
	\tilde{\mathcal{M}}^2_A=\begin{pmatrix}
		\frac{\mathcal{M}^{(0)T}}{\tau^0}+\frac{\mathcal{M}^{(-2)T}}{\tau^2}+\frac{\mathcal{M}^{(-4)}}{\tau^4} & \frac{\mathcal{M}^{(-2)}-\mathcal{M}^{(-2)T}}{\tau^2}-\frac{\mathcal{M}^{(-4)}}{\tau^4}\\ \frac{\mathcal{M}^{(-4)}}{\tau^4} & \frac{\mathcal{M}^{(0)}}{\tau^0}+\frac{\mathcal{M}^{(-2)}}{\tau^2}-\frac{\mathcal{M}^{(-4)}}{\tau^4}
	\end{pmatrix},
\end{equation}
\end{widetext}
where
\begin{align}
	\mathcal{M}^{(0)}\thickspace &= \left(\Omega^{-1}\right)_A \big(\Omega\big)_A,\\
	\mathcal{M}^{(-2)} &= \left(\Omega^{-3}\right)_A\big(\Omega\big)_A-\left(\Omega^{-1}\right)_{A}^2,\\
	\mathcal{M}^{(-4)} &= \left(\Omega^{-3}\right)_A \left(\Omega^{-1}\right)_A-\left(\Omega^{-1}\right)_{A}\left(\Omega^{-3}\right)_{A}.
\end{align}
The eigenvalues of $\tilde{\mathcal{M}}^2_A$, which are the squares of the eigenvalues of $\mathcal{M}_A$, depend only on even powers of $1/\vert\tau\vert$. The eigenvalues of $\mathcal{M}_A$ admit a perturbative expansion of the form 
\begin{equation}\label{eq:series}
\lambda_i=\sum_{k=0}^{\infty}\frac{\lambda^{(2k)}_i}{\tau^{2k}}.
\end{equation}
In the following we use the same notation for similar expansions. 

In the flat-space limit we have the eigenvalue problem 
\begin{equation}
\mathcal{M}^{(0)T}v^{(0)}_i  =\Lambda_i^{(0)} v^{(0)}_i,\qquad \mathcal{M}^{(0)} w^{(0)}_i =\Lambda_i^{(0)} w^{(0)}_i,\label{eq:flat_space}
\end{equation}
where $\Lambda_i^{(0)}=\left(\lambda^{(0)}_{i}\right)^2$. The vectors $v^{(0)}_i$ and $ w^{(0)}_i$ are the left and right eigenvectors of $\mathcal{M}^{(0)}$. The leading corrections to the flat-space eigenvalues $\lambda^{(0)}_{i}$ read
\begin{equation}\label{eq:lambda_2_def}
\lambda^{(2)}_i=\frac{1}{2\lambda^{(0)}_i }\frac{w^{(0)T}_i\mathcal{M}^{(-2)T}v^{(0)}_i}{w^{(0)T}_i v^{(0)}_i}.
\end{equation}
For the whole system we have $\Omega\, \Omega^{-3}=\Omega^{-2}=\Omega^{-1}\Omega^{-1}$ along with $\Omega^{-1}\Omega=I_N$. This allows us to write
\begin{align}
\mathcal{M}^{(0)} &= I_n-\left(\Omega^{-1}\right)_B \big(\Omega\big)_B^T,\label{eq:M_def_0}\\
\mathcal{M}^{(-2)T} &= \left(\Omega^{-1}\right)_B \left(\Omega^{-1}\right)_B^T-\big(\Omega\big)_B\left(\Omega^{-3}\right)_B^T.\label{eq:M_def_2}
\end{align}
The form of Eqs. \eqref{eq:M_def_0} and \eqref{eq:M_def_2} is advantageous because we avoid UV singularities that arise in the continuum limit (when, for instance, we integrate over $y$ in an expression such as Eq. \eqref{eq:M2_int} and $y\rightarrow x$ or  $y\rightarrow x^\prime$).
\section{The Continuum Limit}
\subsection{The Kernels}
We are interested in the eigenvalues $\lambda_i^{(0)}$ and the corrections $\lambda_i^{(2)}$ in the continuum limit. Restoring momentarily the cutoff $\epsilon$ implies that the terms $\lambda_i^{(2k)}$ of Eq. \eqref{eq:series} are actually $\epsilon^{2k}\lambda_i^{(2k)}$. As a result, the kernels $\Omega^{m}\left(x,x^\prime\right)$ corresponding to the matrices $\Omega^{m}$ are defined as
\begin{equation}
\Omega^{m}\left(x,x^\prime\right)=\lim_{\epsilon\rightarrow 0} \frac{\Omega^{m}_{ij}}{\epsilon^m},
\end{equation}
where $x=i\epsilon$ and $x^\prime=j\epsilon$. The $A$-block of the matrix $\Omega^{m}$ corresponds to the kernel $\Omega^{m}\left(x,x^\prime\right)$ with the restriction $x\leq R$ and $x^\prime\leq R$.

The spectral decomposition of the coupling matrix of Eq. \eqref{eq:couplings_matrix} is known and one can take the continuum limit to obtain the eigenfunctions and the corresponding eigenvalues \cite{Katsinis:2024gef}. We find
\begin{equation}\label{eq:powers_Omega}
\Omega^m\left(x,x^\prime\right)=\frac{2}{L}\sum_{k=0}^{\infty} \left(\frac{k\pi}{L}\right)^m \sin\frac{k\pi x}{L}\sin\frac{k\pi x^\prime}{L},
\end{equation}
where $x$ and $x^\prime$ are valued in $\left[0,L\right]$. 

The kernels that are relevant for our calculation can be written in closed form as
\begin{align}
\Omega(x,x^\prime)&=\frac{\pi}{4L^2}\left(\frac{1}{\sin^2\frac{\pi (x+x^\prime)}{2L}}-\frac{1}{\sin^2\frac{\pi (x-x^\prime)}{2L}}\right),\\
\Omega^{-1}(x,x^\prime)&=-\frac{1}{\pi}\ln\left\vert\frac{\sin\frac{\pi\left(x-x^\prime\right)}{2L}}{\sin\frac{\pi\left(x+x^\prime\right)}{2L}}\right\vert,\\
\begin{split}
\Omega^{-3}(x,x^\prime)&=\frac{L^2}{2\pi^3}\bigg[\mathrm{Li}_3\left(e^{-i\frac{\pi}{L}\left(x-x^\prime\right)}\right) \\  &\qquad\qquad-\mathrm{Li}_3\left(e^{-i\frac{\pi}{L}\left(x+x^\prime\right)}\right)\bigg]+\textrm{c.c.}\thickspace.
\end{split}
\end{align}
The regime of interest is $L\gg R$, where $R$ is the radius of the entangling surface. In this regime the kernels assume the form
\begin{align}
\Omega(x,x^\prime)&=\frac{1}{\pi}\left(\frac{1}{(x+x^\prime)^2}-\frac{1}{(x-x^\prime)^2}\right),\\
\Omega^{-1}(x,x^\prime)&=-\frac{1}{\pi}\ln\left\vert\frac{x-x^\prime}{x+x^\prime}\right\vert,\label{eq:Om_m1_inf_size}\\
\begin{split}
\Omega^{-3}(x,x^\prime)&=\frac{1}{\pi}\Bigg[x x^\prime\left(3+2\ln\frac{L}{\pi R}\right)\\&\hspace*{-1cm}+\frac{\left(x-x^\prime\right)^2}{2}\ln \frac{\left\vert x-x^\prime\right\vert}{R}-\frac{\left(x+x^\prime\right)^2}{2}\ln\frac{x+x^\prime}{R}\Bigg].\label{eq:Om_m3_inf_size}
\end{split}
\end{align}

The solution of the eigenvalue problem \eqref{eq:flat_space} is known \cite{Callan:1994py,Katsinis:2024gef}. The right eigenfunctions $f(x;\omega)$, the left eigenfunctions $g(x;\omega)$, and the eigenvalues $\Lambda^{(0)}(\omega)$ of the kernel $\mathcal{M}^{(0)}$ read:
\begin{align}
f(x;\omega)&=\sin\left(\omega u(x)\right),\label{eq:eig_f_zeroth}\\
g(x;\omega)&=\frac{1}{R}\cosh^2\frac{u}{2}\sin\left(\omega u(x)\right),\label{eq:eig_g_zeroth}\\
\Lambda^{(0)}(\omega)&=\coth^2\left(\pi\omega\right),\label{eq:eigv_zeroth}
\end{align}
where $x$ is related to the variable $u$ as
\begin{equation}\label{eq:change_of_var}
x(u)=R \tanh\frac{u}{2},\qquad u(x)=\ln\frac{R+x}{R-x}.
\end{equation}

We need the inner product $v^{(0)T}_i w^{(0)}_i$, which is divergent in the continuum limit. We impose a cutoff at $x=R-\epsilon$, where $\epsilon\ll R$. Thus, we regularize this divergence and at the same time we discretize the spectrum as
\begin{equation}\label{eq:omega_dis}
\omega_k =\frac{ k \pi }{u_{\textrm{max}}}, \qquad u_{\textrm{max}}=\ln \frac{2R}{\epsilon},\qquad k\in \mathbb{N}^*.
\end{equation}
It then follows that the continuum limit of the inner product $w^{(0)T}_i v^{(0)}_i$ is given by
\begin{equation}\label{eq:normalization}
w^{(0)T}_i v^{(0)}_i\rightarrow\frac{1}{2} \int_0^{u_{\textrm{max}}}  du\sin^2\left(\omega_i u\right)= \frac{1}{4}u_\textrm{max}.
\end{equation}

\subsection{The Correction to the Eigenvalues}
We next proceed with the main part of our derivation, which is the calculation of
\begin{multline}
\left\langle M^{(-2)T}\right\rangle\\
=\int_0^R dx \int_0^R dx^\prime f(x;\omega)M^{(-2)T}\left(x,x^\prime\right)g(x^\prime;\omega).
\end{multline}
The kernel $M^{(-2)T}\left(x,x^\prime\right)$, which is the continuum limit of the matrix $M^{(-2)T}$ in the form of Eq. \eqref{eq:M_def_2}, reads
\begin{multline}\label{eq:M2_int}
M^{(-2)T}\left(x,x^\prime\right)=\int^{L}_{\ell} dy \left[\Omega^{-1}(x,y)\Omega^{-1}(y,x^\prime)\right. \\ \left.-\Omega(x,y)\Omega^{-3}(y,x^\prime)\right].
\end{multline}
Taking into account that
\begin{equation}
\Omega^m(x,y)=-\frac{d^2}{dy^2}\Omega^{m-2}(x,y),
\end{equation}
which is a direct consequence of Eq. \eqref{eq:powers_Omega}, we obtain
\begin{multline}
M^{(-2)T}\left(x,x^\prime\right)=\Omega^{-1}(x,R)\left[\frac{\partial}{\partial R}\Omega^{-3}(R,x^\prime)\right]\\-\left[\frac{\partial}{\partial R}\Omega^{-1}(x,R)\right]\Omega^{-3}(R,x^\prime).
\end{multline}
Notice that $M^{(-2)T}\left(x,x^\prime\right)$ is a sum of products of functions of either $x$ or $x^\prime$. As a result, it is finite when $x^\prime\rightarrow x$. Divergences occur only when $x\rightarrow R$ or $x^\prime\rightarrow R$. For $L\gg R$, using Eqs. \eqref{eq:Om_m1_inf_size} and \eqref{eq:Om_m3_inf_size}, we obtain 
\begin{multline}
M^{(-2)T}\left(x,x^\prime\right)=\frac{1}{\pi^2}\left[ x^\prime+\frac{R^2-x^{\prime2}}{2R}\ln\frac{R+x^\prime}{R-x^\prime}\right]\frac{2R x}{R^2-x^2}\\
+\frac{1}{\pi^2}\left[\frac{2R x}{R^2-x^2}+\ln\frac{R+x}{R-x}\right]\left[2x^\prime\left(\ln\frac{L}{\pi R}+1\right)\right. \\ \left.+\left(R-x^\prime\right)\ln\frac{R-x^\prime}{R}-\left(R+x^\prime\right)\ln\frac{R+x^\prime}{R}\right].
\end{multline} 
Substituting the eigenfunctions, given by Eqs. \eqref{eq:eig_f_zeroth} and \eqref{eq:eig_g_zeroth}, and performing the change of variable \eqref{eq:change_of_var}, one obtains $\left\langle M^{(-2)T}\right\rangle$. Making use of the normalization \eqref{eq:normalization}, the corrections of the eigenvalues $\lambda^{(2)}_i$, defined in Eq. \eqref{eq:lambda_2_def}, assume the form 
\begin{multline}\label{eq:lambda_2}
\lambda^{(2)}\left(\omega_i\right)=R^2\frac{\left[I_1\left(\omega_i\right)\left(3+2\ln\frac{L}{\pi R}\right)+I_2\left(\omega_i\right)\right]}{u_\textrm{max} \coth\left(\pi\omega_i\right)}\\\times \left[I_1\left(\omega_i\right)+\frac{1}{2}I_3\left(\omega_i\right)\right],
\end{multline}
where
\begin{align}
I_1\left(\omega\right) &= \frac{1}{\pi}\int_0^{u_{\textrm{max}}} du\sin\left(\omega u\right)\tanh \frac{u}{2},\\
\begin{split}
I_2\left(\omega\right) &= \frac{1}{\pi}\int_0^{u_{\textrm{max}}} du\sin\left(\omega u\right)\bigg(\frac{e^{-u/2}}{\cosh\frac{u}{2}}\ln\frac{e^{-u/2}}{\cosh\frac{u}{2}}\\&\qquad\qquad\qquad\qquad\quad\thickspace-\frac{e^{u/2}}{\cosh\frac{u}{2}}\ln\frac{e^{u/2}}{\cosh\frac{u}{2}}\bigg),
\end{split}\\
I_3\left(\omega\right) &= \frac{1}{\pi}\int_0^{u_{\textrm{max}}} du \, \frac{u\sin\left(\omega u\right)}{\cosh^2\frac{u}{2}}.
\end{align}

We are interested in the $u_{\textrm{max}}\rightarrow\infty$ limit of these integrals, which are calculated as contour integrals. We obtain
\begin{align}
I_1\left(\omega\right) &= \frac{1}{\sinh\left(\pi\omega\right)},\label{eq:integral_1}\\
I_1\left(\omega\right)+\frac{1}{2}I_3\left(\omega\right) &= \pi\omega\frac{\coth\left(\pi\omega\right)}{\sinh\left(\pi\omega\right)}.\label{eq:integral_2}
\end{align}
We do not make use of the explicit expression for $I_2\left(\omega\right)$ in the following, thus we do not present it.
\subsection{The Correction to the Entanglement Entropy}
Following the notation of the expansion \eqref{eq:series}, the first non-vanishing correction to the entanglement entropy of flat space $S^{(0)}$ is given by
\begin{equation}
S^{(2)}=\sum_i \lambda^{(2)}_i\mathrm{arccoth}\,\lambda^{(0)}_i=\sum_i \pi\omega_i \lambda^{(2)}_i.
\end{equation}
Using Eqs. \eqref{eq:lambda_2}, \eqref{eq:integral_1} and \eqref{eq:integral_2}, we obtain
\begin{multline}
S^{(2)}=R^2\frac{\pi}{u_\textrm{max}}\sum_i\left[\left(3+2\ln\frac{L}{\pi R}\right) \frac{\pi\omega_i^2}{\sinh^2\left(\pi\omega_i\right)}\right. \\\left.+\frac{\pi\omega_i I_2\left(\omega_i\right)}{\sinh\left(\pi\omega_i\right)}\right].
\end{multline}
Since $R\gg\epsilon$ the eigenvalues $\omega_i$ become dense and the sum can be approximated by an integral over $\omega$ \cite{Callan:1994py} using Eq. \eqref{eq:omega_dis}, i.e.
\begin{multline}
S^{(2)}=R^2\int_{0}^{\infty}d\omega\left[\left(3+2\ln\frac{L}{\pi R}\right) \frac{\pi\omega^2}{\sinh^2\left(\pi\omega\right)}\right.\\ \left.+\frac{\pi\omega I_2\left(\omega\right)}{\sinh\left(\pi\omega\right)}\right].
\end{multline}
The integral of the first term can be calculated straightforwardly. Regarding the second term, we perform first the integral over $\omega$, using
\begin{equation}
\int_{0}^{\infty}d\omega\frac{\omega\sin\left(\omega u\right)}{\sinh\left(\pi\omega\right)}=\frac{\sinh\frac{u}{2}}{4\cosh^3\frac{u}{2}}
\end{equation}
and then we integrate over $u$, obtaining
\begin{multline}
\int_0^{\infty} du\frac{\sinh\frac{u}{2}}{4\cosh^3\frac{u}{2}}\left(\frac{e^{-u/2}}{\cosh\frac{u}{2}}\ln\frac{e^{-u/2}}{\cosh\frac{u}{2}}\right.\\ \left.-\frac{e^{u/2}}{\cosh\frac{u}{2}}\ln\frac{e^{u/2}}{\cosh\frac{u}{2}}\right)=-\frac{1}{18}-\frac{1}{3}\ln2.
\end{multline}

Putting everything together, the correction to the flat-space entanglement entropy, arising from the $\ell=0$ sector at order $1/\tau^2$, reads
\begin{equation}
S^{(2)}=\left(\frac{1}{3}\ln\frac{L}{2\pi R}+\frac{4}{9}\right)R^2.
\label{1dfinal}
\end{equation}
The higher-$\ell$ sectors give contributions independent of $L$, but with a logarithmic dependence on $\epsilon$ \cite{Boutivas:2024aq}. The result \eqref{1dfinal} determines the $L$-dependence of the total entropy. This means that the precise form of Eq. \eqref{eq:S_IR}, including the numerical coefficient, is
\begin{equation}
S^{\textrm{dS}}_{\textrm{IR}}=\frac{1}{3}\frac{R^2}{\tau^2}\ln\frac{L}{\epsilon}.
\label{finalresult}
\end{equation}

\section{Discussion}
In an accompanying publication \cite{Boutivas:2024aq} we calculate numerically the entanglement entropy for a free field in $(3+1)$-dimensional de Sitter space. The numerical study confirms the appearance of the term \eqref{finalresult} as the leading correction to the flat-space entanglement entropy. The numerical coefficient equal to $1/3$ is reproduced with very high accuracy. It is worth mentioning that the analysis in \cite{Boutivas:2024aq} employs the original discretization scheme of Srednicki \cite{Srednicki:1993im}, which differs slightly from the one we employed here. This confirms that the continuum limit is independent of the details of the discretization. 

The logarithmic term \eqref{finalresult} originates in the sector with vanishing angular momentum, which is equivalent to a $(1+1)$-dimensional theory. Moreover, it results from the lowest frequency modes, whose wavelengths are comparable to the size $L$ of the \emph{overall system}. In flat space, their contribution is suppressed. However, the cosmological expansion induces the strong squeezing of these modes, enhancing their effect on entanglement. This is clearly seen through the qualitative change of the finite-size corrections from equation \eqref{finitesizeflat} to \eqref{finitesizedS}.

It must be pointed out that the consistency of our perturbative calculation requires the term \eqref{finalresult} to be subleading to the flat-space area-law term. We have used an expansion in $1/|\tau|$, but the various orders include powers of the combination $R^2/\tau^2$. In terms of the physical quantities $R_p=a R$, $\epsilon_p=a \epsilon$ and $H=-1/(a\tau)$, this limits us to subhorizon entangling radii $\epsilon_p\ll R_p \ll 1/H$. However, the overall system size $L_p=a L$ can exceed the Hubble radius, as long as it is not exponentially large to invalidate the expansion.

The presence of a novel term that depends logarithmically on the size of the \emph{overall system} is intriguing. This term is not dominant within the region of validity of the perturbative expansion we employed, and thus it does not determine the scaling properties of the entanglement entropy. It should be noted, however, that a presence of a term with a logarithmic dependence on $L$ is likely even in late-time or superhorizon regimes over which we do not have analytical control at present. The reason is that such a term is incorporated in the kernel $\Omega^{-3}$ given by \eqref{eq:Om_m3_inf_size}, which appears in the full matrix \eqref{eq:matrix_M_cal}. 

Even though it is unclear if the term \eqref{finalresult} can become dominant in some regime, its importance cannot be overstated: it represents the leading effect of the cosmological expansion for subhorizon entangling radii, and also provides a probe of regions beyond the horizon, which are otherwise inaccessible. A relevant question concerns the identification of the parameter $L$. For a finite universe, this quantity is fixed by its total size, which can be so large that its presence cannot be deduced through curvature effects at subhorizon distances. But even for a spatially flat infinite universe, there exists a bound on $L$, set by the duration of inflation. This can be seen by considering the function $b(\tau)$ that quantifies the effect of squeezing and determines the wave function \eqref{solschrod} of the field modes. The assumption of a Bunch-Davies vacuum in exact de Sitter space fixes this function to have the form $b(\tau)=\sqrt{1+1/(\omega^2_0 \tau^2)}$, so that $b(\tau)\to 1$ and $b'(\tau)\to 0$ for $\tau\to -\infty$. If, however, we assume that inflation started at some finite time $\tau_0$, with no prior squeezing of the wave function, these conditions must be imposed at $\tau_0$. It can be checked that $b(\tau)$ is not modified considerably for modes with $\omega_0\tau_0\gg 1$. Such modes start subhorizon and accumulate all the squeezing induced by the subsequent expansion. However, for modes with $\omega_0\tau_0\ll 1$, the function $b(\tau)$ is suppressed by $\sim \omega_0\tau_0$ at late times, relatively to the exact de Sitter scenario. Such modes are already superhorizon in the beginning of inflation, so that their total squeezing is reduced. It can then be checked that these modes do not contribute to the kernel $\Omega^{-3}$ that induces the logarithmic term. Thus $L$ can be identified with the wavelength of the first mode that exited the horizon in the beginning of inflation, which may extend far beyond the present-day horizon.  

The dependence of the entanglement entropy on the size of the overall system implies that the degrees of freedom that lie beyond the cosmological horizon affect its interior. Classically, this is not permitted. However, the nature of quantum entanglement allows for this possibility. At the conceptual level this effect is interesting because of its strong similarity to the EPR paradox. A possibility to test the effect is provided by analogue gravity experiments with ultra-cold atoms, which can simulate field evolution in an expanding background \cite{Eckel:2017uqx,Wittemer:2019agm,Viermann:2022wgw,Tolosa-Simeon:2022umw}. Similar experiments have measured entanglement entropy, or related quantities, in a static setting \cite{Islam:2015mom,Brydges:2019wut,Tajik:2022ycs}. An experiment that combines elements from both these lines of research could provide a quantitative check of the dependence of the entropy on the size of the overall system.

\bibliography{spooky_analytic}

\end{document}